\documentclass[a4paper,11pt]{article}
\usepackage{pos}

\title{Integration techniques for worldline integrals}

\author[a]{Victor M. Banda Guzmán}
\author[b]{James P. Edwards}
\author[c]{C. Moctezuma Mata Zamora}
\author[d,e]{Luis A. Rodriguez Chacón}
\author*[f]{Christian Schubert}

\affiliation[a]{Universidad Politécnica de San Luis Potosí, Urbano Villalón 500, Colonia La Ladrillera, C.P. 78363, San Luis Potosí, San Luis Potosí, Mexico}
\affiliation[b]{Centre for Mathematical Sciences, University of Plymouth, Plymouth, PL4 8AA, UK}
\affiliation[c]{Instituto de Física y Matemáticas, Universidad Michoacana de San Nicolás de Hidalgo, Edificio C-3,Apdo. Postal 2-82, 8060 Morelia, Michoacán, Mexico}
\affiliation[d]{Computation-based Science and Technology Research Center, The Cyprus Institute,
20 Konstantinou Kavafi Street, 2121, Aglantzia, Nicosia, Cyprus}
\affiliation[e]{Department of Physics and Earth Science, University of Ferrara \& INFN, Via Saragat 1, 44122 Ferrara, Italy}
\affiliation[f]{Facultad de Ciencias Físico-Matemáticas, Universidad Michoacana de San Nicolás de Hidalgo, Avenida Francisco J. Mújica, 58060 Morelia, Michoacán, Mexico} 

\emailAdd{christianschubert137@gmail.com}

\abstract{The worldline formalism allows one to obtain compact integral representations combining the information of large numbers of Feynman diagrams. However, their analytic calculation leads to a non-standard integration problem for which existing mathematical algorithms are of little help. Here I will summarize the state-of-the-art of worldline integration focusing on examples from QED in vacuum and in constant external fields.
}

\FullConference{17th International Symposium on Radiative Corrections: Applications of Quantum Field Theory to Phenomenology (RADCOR2025)\\
5-10 October 2025\\
Puri, India\\}

\begin{document}

\def\nonu{\\}
\newcommand{\intT}{\int_0^\infty \dfrac{dT}{T}\, T^{4-\frac{D}{2}}\,e^{-m^2 T}\,
\int_{(4)} 
\e^{T\,\Lambda} \;}

\newcommand{\g}[1]{G_{ #1}}
\newcommand{\dg}[1]{\dot{G}_{ #1}}
\newcommand{\ddg}[1]{\ddot{G}_{ #1}}
\newcommand{\dilog}[1]{\text{Li}_2\left(#1\right)}
\newcommand{\Log}[1]{\log\left(#1\right)}
\newcommand{\hs}{\hat{s}}
\newcommand{\hT}{\hat{t}}
\newcommand{\hu}{\hat{u}}
\newcommand{\ha}{\hat{a}}
\newcommand{\hb}{\hat{b}}
\newcommand{\hc}{\hat{c}}
\newcommand{\e}{\text{e}}
\newcommand{\Arg}{\text{Arg}}
\newcommand{\bet}[1]{\beta_{\hat{#1}}}

\newcommand{\lnbeta}[1]{\ln\left( \dfrac{\beta_{#1}-1}{\beta_{#1}+1} \right)}

\newcommand{\bifmm}[3]{\bi{ -\dfrac{\beta_{#1}}{2},-\dfrac{\beta_{#2#3}}{2} }}
\newcommand{\bifmp}[3]{\bi{ -\dfrac{\beta_{#1}}{2}, \dfrac{\beta_{#2#3}}{2} }}
\newcommand{\bifpp}[3]{\bi{ \dfrac{\beta_{#1}}{2}, \dfrac{\beta_{#2#3}}{2} }}
\newcommand{\bifpm}[3]{\bi{ \dfrac{\beta_{#1}}{2}, -\dfrac{\beta_{#2#3}}{2} }}

\def\ddel{{}^\bullet\! \Delta}
\def\deld{\Delta^{\hskip -.5mm \bullet}}
\def\dddel{{}^{\bullet \bullet} \! \Delta}
\def\ddeld{{}^{\bullet}\! \Delta^{\hskip -.5mm \bullet}}
\def\deldd{\Delta^{\hskip -.5mm \bullet \bullet}}
\def\epsk#1#2{\varepsilon_{#1}\cdot k_{#2}}
\def\epseps#1#2{\varepsilon_{#1}\cdot\varepsilon_{#2}}

\newcommand{\Ascr}{\mathscr{A}}
\newcommand{\Dscr}{\mathscr{D}}
\newcommand{\Mscr}{\mathscr{M}}
\newcommand{\Wscr}{\mathscr{W}}
\newcommand{\OWscr}{\mathscr{OW}}

\newcommand{\Acal}{\mathcal{A}}
\newcommand{\Gcal}{\mathcal{G}}
\newcommand{\Dcal}{\mathcal{D}}
\newcommand{\Mcal}{\mathcal{M}}

\newcommand{\jhat}[1]{\hspace{0.3em}\widehat{\hspace{-0.4em}#1\hspace{-0.4em}}\hspace{0.4em}}
\newcommand{\bone}{1\!\!1}

%
\def\cosech{\rm cosech}
\def\sech{\rm sech}
\def\coth{\rm coth}
\def\tanh{\rm tanh}
\def\half{{1\over 2}}
\def\third{{1\over3}}
\def\fourth{{1\over4}}
\def\fifth{{1\over5}}
\def\sixth{{1\over6}}
\def\seventh{{1\over7}}
\def\eigth{{1\over8}}
\def\ninth{{1\over9}}
\def\tenth{{1\over10}}
\def\bN{\mathop{\bf N}}
\def\R{{\rm I\!R}}
\def\Eins{{\mathchoice {\rm 1\mskip-4mu l} {\rm 1\mskip-4mu l}
{\rm 1\mskip-4.5mu l} {\rm 1\mskip-5mu l}}}
\def\Z{{\mathchoice {\hbox{$\sf\textstyle Z\kern-0.4em Z$}}
{\hbox{$\sf\textstyle Z\kern-0.4em Z$}}
{\hbox{$\sf\scriptstyle Z\kern-0.3em Z$}}
{\hbox{$\sf\scriptscriptstyle Z\kern-0.2em Z$}}}}
\def\abs#1{\left| #1\right|}
\def\com#1#2{
        \left[#1, #2\right]}
\def\square{\kern1pt\vbox{\hrule height 1.2pt\hbox{\vrule width 1.2pt
   \hskip 3pt\vbox{\vskip 6pt}\hskip 3pt\vrule width 0.6pt}
   \hrule height 0.6pt}\kern1pt}
      \def\boxop{{\raise-.25ex\hbox{\square}}}
\def\contract{\makebox[1.2em][c]{
        \mbox{\rule{.6em}{.01truein}\rule{.01truein}{.6em}}}}
\def\ltap{\ \raisebox{-.4ex}{\rlap{$\sim$}} \raisebox{.4ex}{$<$}\ }
\def\gtap{\ \raisebox{-.4ex}{\rlap{$\sim$}} \raisebox{.4ex}{$>$}\ }
\def\mn{{\mu\nu}}
\def\rs{{\rho\sigma}}
\newcommand{\Det}{{\rm Det}}
\def\Tr{{\rm Tr}\,}
\def\tr{{\rm tr}\,}
\def\sumij{\sum_{i<j}}
\def\e{\,{\rm e}}
\def\non{\\}
\def\br{{\bf r}}
\def\bp{{\bf p}}
\def\bx{{\bf x}}
\def\by{{\bf y}}
\def\brhat{{\bf \hat r}}
\def\bv{{\bf v}}
\def\ba{{\bf a}}
\def\bE{{\bf E}}
\def\bB{{\bf B}}
\def\bA{{\bf A}}
\def\pa{\partial}
\def\dA{\partial^2}
\def\ddx{{d\over dx}}
\def\ddt{{d\over dt}}
\def\der#1#2{{d #1\over d#2}}
\def\lie{\hbox{\it \$}} 
\def\partder#1#2{{\partial #1\over\partial #2}}
\def\secder#1#2#3{{\partial^2 #1\over\partial #2 \partial #3}}
\def\kinq{{1\over 4}\dot q^2}
\def\kinb{{1\over 4}\dot x^2}
%
\def\bef{}
\def\ef{}
\def\be{\begin{equation}}
\def\ee{\end{equation}\noindent}
\def\bear{\begin{eqnarray}}
\def\ear{\end{eqnarray}\noindent}
\def\bec{\begin{equation}}
\def\eec{\end{equation}\noindent}
\def\bearc{\begin{eqnarray}}
\def\earc{\end{eqnarray}\noindent}
\def\benn{\begin{enumerate}}
\def\enn{\end{enumerate}}
\def\veject{\vfill\eject}
\def\ven{\vfill\eject\noindent}
%
\def\eq#1{{eq. (\ref{#1})}}
\def\eqs#1#2{{eqs. (\ref{#1}) -- (\ref{#2})}}
%
\def\totint{\int_{-\infty}^{\infty}}
\def\posint{\int_0^{\infty}}
\def\negint{\int_{-\infty}^0}
\def\pint{{\dps\int}{dp_i\over {(2\pi)}^d}}
%
\newcommand{\GeV}{\mbox{GeV}}
\def\FFdual{F\cdot\tilde F}
\def\bra#1{\langle #1 |}
\def\ket#1{| #1 \rangle}
\def\braket#1#2{\langle {#1} \mid {#2} \rangle}
\def\vev#1{\langle #1 \rangle}
\def\rightvac{\mid 0\rangle}
\def\leftvac{\langle 0\mid}
\def\ihbar{{i\over\hbar}}
\def\ge{\hbox{$\gamma_1$}}
\def\gz{\hbox{$\gamma_2$}}
\def\gd{\hbox{$\gamma_3$}}
\def\go{\hbox{$\gamma_1$}}
\def\gt{\hbox{\$\gamma_2$}}
\def\gth{\hbox{$\gamma_3$}} 
\def\gf{\hbox{$\gamma_5\;$}}
\def\slash#1{#1\!\!\!\raise.15ex\hbox {/}}
\newcommand{\slD}{\,\raise.15ex\hbox{$/$}\kern-.27em\hbox{$\!\!\!D$}}
\newcommand{\slpartial}{\raise.15ex\hbox{$/$}\kern-.57em\hbox{$\partial$}}

\newcommand{\PP}{\cal P}
\newcommand{\G}{{\cal G}}
\newcommand{\nc}{\newcommand}
\newcommand{\Fkala}{F_{\kappak_i\cdot k_j}}
\newcommand{\Fkanu}{F_{\kappa\nu}}
\newcommand{\Flaka}{F_{k_i\cdot k_j\kappa}}
\newcommand{\Flamu}{F_{k_i\cdot k_j\mu}}
\newcommand{\Fmunu}{F_{\mu\nu}}
\newcommand{\Fnumu}{F_{\nu\mu}}
\newcommand{\Fnuka}{F_{\nu\kappa}}
\newcommand{\Fmuka}{F_{\mu\kappa}}
\newcommand{\Fkalamu}{F_{\kappak_i\cdot k_j\mu}}
\newcommand{\Flamunu}{F_{k_i\cdot k_j\mu\nu}}
\newcommand{\Flanumu}{F_{k_i\cdot k_j\nu\mu}}
\newcommand{\Fkamula}{F_{\kappa\muk_i\cdot k_j}}
\newcommand{\Fkanumu}{F_{\kappa\nu\mu}}
\newcommand{\Fmulaka}{F_{\muk_i\cdot k_j\kappa}}
\newcommand{\Fmulanu}{F_{\muk_i\cdot k_j\nu}}
\newcommand{\Fmunuka}{F_{\mu\nu\kappa}}
\newcommand{\Fkalamunu}{F_{\kappak_i\cdot k_j\mu\nu}}
\newcommand{\Flakanumu}{F_{k_i\cdot k_j\kappa\nu\mu}}

\nc{\spa}[3]{\left\langle#1\,#3\right\rangle}
\nc{\spb}[3]{\left[#1\,#3\right]}
\nc{\ksl}{\not{\hbox{\kern-2.3pt $k$}}}
\nc{\hf}{\textstyle{1\over2}}
\nc{\pol}{\varepsilon}
\nc{\tq}{{\tilde q}}
\nc{\esl}{\not{\hbox{\kern-2.3pt $\pol$}}}
\newcommand{\cL}{\cal L}
\newcommand{\D}{\cal D}
\newcommand{\Dhalf}{{D\over 2}}
\def\eps{\epsilon}
\def\epshalf{{\epsilon\over 2}}
\def\lag{( -\partial^2 + V)}
\def\freeexp{{\rm e}^{-\int_0^Td\tau {1\over 4}\dot x^2}}
\def\kinb{{1\over 4}\dot x^2}
\def\kinf{{1\over 2}\psi\dot\psi}
\def\expk{{\rm exp}\biggl[\,\sum_{i<j=1}^4 G_{Bij}k_i\cdot k_j\biggr]}
\def\expp{{\rm exp}\biggl[\,\sum_{i<j=1}^4 G_{Bij}p_i\cdot p_j\biggr]}
\def\expshort{{\e}^{\half G_{Bij}k_i\cdot k_j}}
\def\expabb{{\e}^{(\cdot )}}
\def\epseps#1#2{\varepsilon_{#1}\cdot \varepsilon_{#2}}
\def\epsk#1#2{\varepsilon_{#1}\cdot k_{#2}}
\def\kk#1#2{k_{#1}\cdot k_{#2}}
\def\G#1#2{G_{B#1#2}}
\def\Gp#1#2{{\dot G_{B#1#2}}}
\def\GF#1#2{G_{F#1#2}}
\def\Dab{{(x_a-x_b)}}
\def\Dsq{{({(x_a-x_b)}^2)}}
\def\PITD{{(4\pi T)}^{-{D\over 2}}}
\def\4piTD{{(4\pi T)}^{-{D\over 2}}}
\def\4piT4{{(4\pi T)}^{-2}}
\def\TintmD{{\dps\int_{0}^{\infty}}{dT\over T}\,e^{-m^2T}
    {(4\pi T)}^{-{D\over 2}}}
\def\Tintm4{{\dps\int_{0}^{\infty}}{dT\over T}\,e^{-m^2T}
    {(4\pi T)}^{-2}}
\def\Tintm{{\dps\int_{0}^{\infty}}{dT\over T}\,e^{-m^2T}}
\def\Tint{{\dps\int_{0}^{\infty}}{dT\over T}}
\def\np{n_{+}}
\def\nm{n_{-}}
\def\Np{N_{+}}
\def\Nm{N_{-}}
\newcommand{\slG}{{{\dot G}\!\!\!\! \raise.15ex\hbox {/}}}
\newcommand{\Gd}{{\dot G}}
\newcommand{\Gund}{{\underline{\dot G}}}
\newcommand{\Gdd}{{\ddot G}}
\def\GBd12{{\dot G}_{B12}}
\def\Dx{\dps\int{\cal D}x}
\def\Dy{\dps\int{\cal D}y}
\def\Dpsi{\dps\int{\cal D}\psi}
\def\dint#1{\int\!\!\!\!\!\int\limits_{\!\!#1}}
\def\ddtau{{d\over d\tau}}
\def\ie{\hbox{$\textstyle{\int_1}$}}
\def\iz{\hbox{$\textstyle{\int_2}$}}
\def\id{\hbox{$\textstyle{\int_3}$}}
\def\ldop{\hbox{$\lbrace\mskip -4.5mu\mid$}}
\def\rdop{\hbox{$\mid\mskip -4.3mu\rbrace$}}
%
\newcommand{\1}{{\'\i}}
\newcommand{\no}{\noindent}
\def\non{}
\def\dps{\displaystyle}
\def\sy{\scriptscriptstyle}
\def\sy{\scriptscriptstyle}

\maketitle

\section{Worldline representation of the QED S-matrix}

Already at the dawn of modern quantum field theory, Feynman \cite{feynman1950,feynman1951} realized that
the perturbative QED S-matrix can be written in terms of first-quantized relativistic particle path integrals, representing the electrons and positrons, interconnected by photons in all possible ways.
And this is actually a more compact representation than the usual one in terms of Feynman diagrams. 

\begin{figure}[htbp]
\begin{center}
 \includegraphics[width=0.45\textwidth]{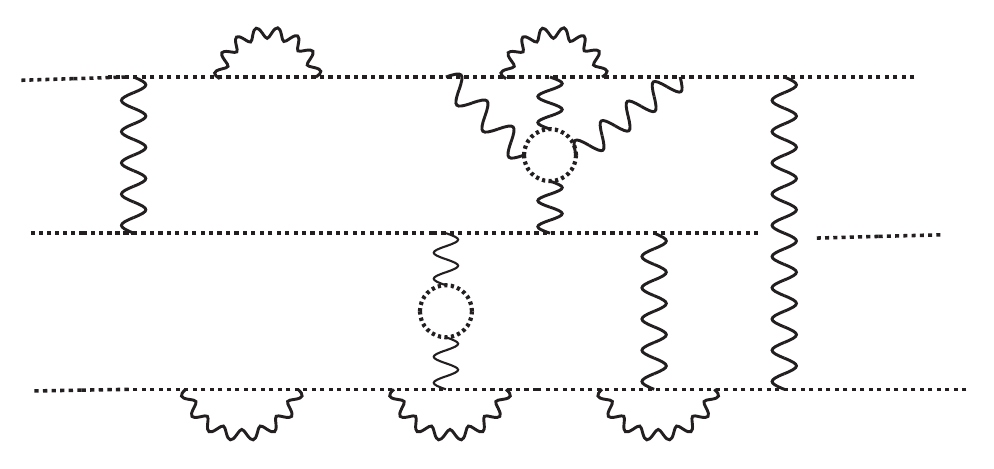}
\caption{{A typical multi-loop Feynman digram in QED.}}
\label{fig-QEDSmatrix}
\end{center}
\end{figure}

For example, the single Feynman diagram shown in Fig. \ref{fig-QEDSmatrix} in Feynman's worldline representation corresponds to a sum of several thousand Feynman diagrams, since 
in this representation photons are free to cross each other, so that it allows one to combine all diagrams differing only by the ordering of some photons along an electron line or loop.
This fact leads to a very significant reduction in the number of integrals generated in perturbation theory at higher loop orders; for example, the 12672 encountered in the Feynman-diagram calculation of the
five-loop contribution to the QED $g-2$ factor in this approach can be reduced to a total of 32 integrals\footnote{We thank N. Ahmadiniaz for this information.}.
It is also the main motivation for much of what we are going to do in the following. 

The derivation of the worldline representation nowadays usually starts with the following well-known formula for the one-loop effective action induced by a scalar loop
for an external Maxwell field $A^\mu(x)$,
\bear
\Gamma [A] &=&- {\rm \Tr}{\rm ln} \Bigl\lbrack -(\partial + ie A)^2+m^2\Bigr\rbrack 
= \int_0^{\infty} \frac{dT}{T} \,{\rm Tr} \, {\rm exp}\Bigl\lbrack - T (  -(\partial + ie A)^2+m^2)\Bigr\rbrack \nonumber \\
&=&
\int_0^{\infty}
\frac{dT}{T}\,
\e^{-m^2T}
\int_{x(0)=x(T)}
{\cal D}x(\tau)\,
e^{-\int_0^T d\tau \bigl(\kinb +ie\dot x\cdot A(x(\tau))\bigr)}
\, .
\label{scalarloop}
\ear
Here $m$ and $T$ are the mass and the proper-time of the loop scalar, and the path integral has to be computed, at fixed $T$,
over all closed loops in euclidean spacetime with periodicity $T$. 

Expanding the field in $N$ plane waves,
$A^{\mu}(x(\tau)) = \sum_{i=1}^N \,\varepsilon_i^{\mu}\e^{ik_i\cdot x(\tau)}$,
and picking out the terms that involve each photon once one gets 
a path-integral representation of the one-loop $N$ - photon amplitudes.

In Feynman's original approach \cite{feynman1951} the electron spin was added by inserting, under the path integral,
the ``Feynman spin factor''  ${\rm Spin}[x,A]$,
\bear
{\rm Spin}[x,A] = {\rm tr}_{\Gamma} {\cal P}
\exp\Biggl[{
\frac{i}{4}
e\,[\gamma^{\mu},\gamma^{\nu}]
\int_0^Td\tau F_{\mu\nu}(x(\tau))}\Biggr]
\label{defSpin}
\ear
where $\cal P$ denotes path ordering.
While this spin factor is still used today for numerical calculations of the worldline path integral,
for analytical purposes nowadays one usually prefers Fradkin's representation of the
spin factor in terms of a Grassmann path integral \cite{fradkin66,fragit91}, 

\bear
{\rm Spin}[x,A] 
=
\int {\cal D}\psi(\tau)
\,
\exp 
\Biggl\lbrack
-\int_0^Td\tau
\Biggl(
\half \psi\cdot \dot\psi -ie \psi^{\mu}F_{\mn}(x(\tau))\psi^{\nu}
\Biggr)
\Biggr\rbrack
\, .
\non
\ear
Here $\psi^\mu(\tau)$ is a Grassmann-valued Lorentz vector fulfilling antiperiodic boundary conditions
in proper time, $\psi^\mu(T)= - \psi^\mu(0)$.
An important point of this Grassmann approach is that it allows one to replace the path-ordered exponential by an ordinary one.

For the scalar case, it is straightforward to generalize the above to a path-integral representation of the scalar propagator
dressed with an arbitrary number of photons \cite{dashsu}. To the contrary, in spinor QED the open-line
case is much more subtle \cite{feynman1951,fragit91,18}, and 
a worldline representation of the photon-dressed electron propagator suitable for state-of-the-art calculations was constructed
only recently,  based on the second-order representation of the Dirac propagator in a Maxwell background and the 
``symbol map'' \cite{130,131}.

\section{String-inspired treatment of the worldline path integral}

After the specialization to the $N$-photon background, all the worldline path integrals mentioned above are gaussian,
which suggests their evaluation through Wick contraction using appropriate worldline Green's functions.
This method is sometimes called ``string-inspired'' since it is analogous to the usual way of calculating the first-quantized
Polyakov path integral in string theory. For the case of the QED $N$-photon amplitudes, this was apparently first
carried through in Polyakov's book \cite{polbook} (see also \cite{strassler,41}).
For scalar QED, it is straightforward to arrive at the following master formula for the 
one-loop $N$-photon amplitudes with momenta $k_i$ and polarizations $\varepsilon_i$,
\begin{eqnarray}
\Gamma[\lbrace k_i,\varepsilon_i\rbrace]
&=&
{(-ie)}^N
{\dps\int_{0}^{\infty}}{dT\over T}
{(4\pi T)}^{-{D\over 2}}
e^{-m^2T}
\prod_{i=1}^N \int_0^T 
d\tau_i
\nonumber\\
&&\hspace{-8pt}
\times
\exp\biggl\lbrace\sum_{i,j=1}^N 
\bigl\lbrack \half G_{ij} k_i\cdot k_j
-i\dot G_{ij}\varepsilon_i\cdot k_j 
+\half\ddot G_{ij}\varepsilon_i\cdot\varepsilon_j
\bigr\rbrack\biggr\rbrace
\mid_{{\rm lin}(\varepsilon_1,\ldots,\varepsilon_N)}
\label{qedmaster}
\end{eqnarray}
\no
where $\tau_i$  parametrizes the position of photon $i$  along the loop,
and we have introduced the worldline Green's function 
$G(\tau_1,\tau_2)$,
\bear
G(\tau_1,\tau_2) &=& \vert \tau_1 -\tau_2\vert -{\bigl(\tau_1 -\tau_2\bigr)^2\over T}\label{defG}
\ear
together with its derivatives 
$\dot G(\tau_1,\tau_2) ={\rm sgn}(\tau_1 - \tau_2)- 2 (\tau_1 - \tau_2)/T, \quad
\ddot G(\tau_1,\tau_2)= 2 {\delta}(\tau_1 - \tau_2) - 2/T\non$.

$D$ is the spacetime dimension, and the notation $\mid_{{\rm lin}(\varepsilon_1,\ldots,\varepsilon_N)}$ means projection on the terms linear in
each polarization vector.

A similar master formula can be written for the spinor loop \cite{polbook,41} using the ``worldline superfield'' $X^\mu = x^\mu +\sqrt{2}\theta \psi^\mu$
and the ``super worldline Green's function''
$\hat G_{12} = G_{12} + \theta_1\theta_2G_{F12}$
involving the ``fermionic worldline Green's function'' $G_F(\tau,\tau') = {\rm sgn}(\tau-\tau')$ and Grassmann parameters $\theta_i$.

\section{Inclusion of a constant external field}
\label{sec:external}

It is also straightforward to further generalize these master formulas to include the effect of an arbitrary constant external field \cite{shaisultanov,18}.
This requires changing the worldline Green's functions $ G, G_F$ to field-dependent ones $ {\cal G}_B, {\cal G}_F$, and also supplying
a global field-dependent determinant factor that, by itself, just reproduces the well-known Weisskopf/Euler-Heisenberg Lagrangians.

With this notation, and the further definition ${\cal Z}_\mn \equiv eF_\mn T$,  the scalar QED master formula \eqref{qedmaster} generalizes to the following representation of the $N$-photon amplitudes
in a constant field,
\begin{eqnarray}
&&\Gamma_{\rm scal}
(k_1,\varepsilon_1;\ldots;k_N,\varepsilon_N\vert F)
=
{(-ie)}^N
{\dps\int_{0}^{\infty}}{dT\over T}
{(4\pi T)}^{-{D\over 2}}
e^{-m^2T}
{\rm det}^{{1\over 2}}
\biggl[
\frac{\cal Z}{{\rm sin}{\cal Z}}
\biggr]
\prod_{i=1}^N \int_0^T 
d\tau_i
\nonumber\\
&&\hspace{45pt}\times
\exp\biggl\lbrace\sum_{i,j=1}^N 
\Bigl\lbrack \half k_i\cdot {\cal G}_{Bij}\cdot  k_j
-i\varepsilon_i\cdot\dot{\cal G}_{Bij}\cdot k_j
+\half
\varepsilon_i\cdot\ddot {\cal G}_{Bij}\cdot\varepsilon_j
\Bigr\rbrack\biggr\rbrace
\Big\vert_{\varepsilon_1\varepsilon_2\cdots \varepsilon_N}\,. 
\label{masterF}
\end{eqnarray}
\no
The master formula \eqref{masterF} and its spinor QED generalization have already shown their usefulness in a number
of calculations. This includes the vacuum polarisation tensor in the field \cite{ditsha,40}, magnetic photon splitting \cite{17},
and the low-energy limit of the $N$-photon amplitudes in a constant field \cite{156,lopezlopez,ahcoed}.
It has also been generalized to higher loops \cite{18,51}, to the inclusion of external gravitons \cite{61,71,161}, and to plane-wave backgrounds \cite{141,ceir}. 

For use below, let us write down here the derivative of the bosonic constant-field Green's function for the case of a purely magnetic field:
\bear
\dot{\cal G}_{B}(\tau_1,\tau_2)
=\dot G_{12}\,{g_-}+ S_{B12}(z)g_+ -A_{B12}(z) i{r_+}
\label{defcalGB}
\ear
where $z = eBT$, 

\bear
g_+\equiv
\left(
\begin{array}{*{4}{c}}
1&0&0&0\\
0&1&0&0\\
0&0&0&0\\
0&0&0&0
\end{array}
\right),\qquad
g_-\equiv
\left(
\begin{array}{*{4}{c}}
0&0&0&0\\
0&0&0&0\\
0&0&1&0\\
0&0&0&1
\end{array}
\right),\nonumber\\
\label{app-gd-gmat}
\ear
\vspace{-30pt}
\begin{equation}
r_+ \equiv
\left(
\begin{array}{*{4}{c}}
0&1&0&0\\
-1&0&0&0\\
0&0&0&0\\
0&0&0&0
\end{array}
\right),\qquad
r_- \equiv
\left(
\begin{array}{*{4}{c}}
0&0&0&0\\
0&0&0&0\\
0&0&0&1\\
0&0&-1&0
\end{array}
\right),
\nonumber
\end{equation}
and with
\bear
S_{B12}(z) &=&
{\sinh(z\,\dot G_{12})\over \sinh z} 
\, ,
\\
A_{B12}(z) &=&
{\cosh(z \,\dot G_{12})\over 
\sinh z}-{1\over z} \;.
\ear

\section{Generalizations}

Shortly after the work of Polyakov, Bern and Kosower \cite{berkosNPB} derived a master formula analogous to \eqref{qedmaster} for the scalar-loop contribution to the
(colour-ordered) one-loop QCD $N$-gluon amplitudes:
\bear
\Gamma^{a_{1}\dots a_{N}}[k_{1},\varepsilon_{1};\dots;k_{N},\varepsilon_{N}]
 &=&(-ig)^{N}\mbox{tr}(T^{a_{1}}\dots T^{a_{N}})
 \int_{0}^{\infty} dT(4\pi T)^{-D/2}e^{-m^2 T}\nonumber\\
 &&\hspace{-140pt} \times\int_{0}^{T}d\tau_{1}\dots\int_0^{\tau_{N-2}}d\tau_{N-1}
 \exp\Bigg\{\sum_{i,j=1}^N\left[\frac{1}{2}
  G_{ij}k_{i}\cdot k_{j}
-i\dot{G}_{ij}\varepsilon_{i}\cdot k_{j}+\frac{1}{2}\ddot{G}_{ij}\varepsilon_{i}\cdot\varepsilon_{j}\right]\Bigg\}
\Biggl\vert_{\rm lin (\varepsilon_1 \ldots \varepsilon_N)} 
\, .
\nonumber\\
\label{qcdmaster}
\ear 
Contrary to the abelian case, the gluons now appear in a fixed ordering along the loop. The $T^{a_i}$ are generators of the colour group in the representation of the loop particle. 

Bern and Kosower moreover used string theory to derive a set of ``replacement rules'', that allow one to derive from this scalar-loop master formula also representations of the fermion and gluon-loop contributions
to the $N$-gluon amplitudes, and ``pinch rules'' to include the one-particle reducible ones, which also exist in the non-abelian case. 

See \cite{158,bafemi} for the generalization to other Standard Model interactions. 

\section{Advantages of the worldline representation in QED}

Before embarking on sample calculations, let us summarize the advantages that one can hope to find in QED when
using the worldline formalism instead of the more standard Feynman diagram techniques:

\benn

\item
Compact integral representations with manifest Bose symmetry between the external photons. 

\item
Avoidance of Dirac algebra work. 

\item
Unification of scalar and spinor QED calculations. 

\item
Fermion lines and loops are treated as a whole rather than segmented into individual propagators. This is particularly
important in the presence of external fields where each propagator has already a complicated structure. 

\item
The worldline representation gives already the complete amplitudes, without the need of summing over crossed diagrams. 

\enn

\section{The fundamental problem of worldline integration}

The last point may not seem particularly relevant at the one-loop level, but becomes important at higher loop levels.
All the master formulas mentioned above are valid off-shell, and thus, like the original path integrals, can still be used to 
construct higher-loop amplitudes. The property of the formalism of combining all diagrams that differ only
by a different ordering of the photon legs along a given fermion line or loop then allows one to obtain integral representations
combining the information of Feynman diagrams of different topologies, different Symanzik polynomials, 
and often even of different numbers of independent Feynman-Schwinger parameter integrations \cite{18,15,41}. 

However, this unification of diagrams comes at a cost: it is made possible by the appearance of absolute values and sign functions in the
Green's functions $G,\dot G$ and $G_F$. In a numerical calculation those do not normally cause troubles, but for analytical purposes usually one would break such integrals
into ordered sectors, which in the present context would essentially mean to undo the unifying effect of the worldline representation. Thus it is essential
to develop methods for performing such integrals ``in one go'', which was called the ``fundamental problem of worldline integration'' in \cite{135}. 
Unfortunately, from a mathematical point of view here we enter essentially uncharted territory. 

For some concrete examples, let us return to the one-loop QED $N$-photon amplitudes. Considering that the $\ddot G_{ij}$'s can always be removed by
suitable IBP's \cite{berkosNPB}, the $G_{Fij}$'s always be eliminated by the identity
$G_{Fij}G_{Fjk}G_{Fki} = - (\dot G_{ij} + \dot G_{jk} + \dot G_{ki})  $,
and that the integration variables $\tau_i$ can be rescaled to run from zero to one, 
the most general integral to be computed here is of the form
\bear
\int_0^1du_1du_2 \cdots du_N \, {\rm Pol} (\dot G_{ij}) \,\e^{\sum_{i<j=1}^N \lambda_{ij}^2 G_{ij} }
\ear
with arbitrary $ N$ and polynomial ${\rm Pol}(\dot G_{ij})$, where now

\bear
G_{ij} = |u_i-u_j| - (u_i-u_j)^2, \quad
\dot G_{ij} = {\rm sgn}(u_i-u_j) - 2(u_i-u_j)
\, .
\ear
The challenge is to develop integration techniques and build tables of formulas that would allow one to perform such ``circular'' integrals without ever decomposing the integrand into ordered sectors.

\section{Circular integrals: the polynomial case}

At the polynomial level, the fundamental problem has been completely solved
through the following master formula \cite{135},
that allows one to integrate an arbitrary monomial in the $\dot G_{ij}$'s in any
of the $u_i$ variables, and obtain the result as a polynomial in the remaining 
$\dot G_{ij}$'s:
\bear
\int_0^1 du\,
\dot G(u,u_1)^{k_1}
\dot G(u,u_2)^{k_2}
\cdots
\dot G(u,u_n)^{k_n}
&=&
{1\over 2n}
\sum_{i=1}^n\,
\prod_{j\ne i}
\sum_{l_j=0}^{k_j}
{k_j\choose l_j}
\dot G_{ij}^{k_j-l_j}
\sum_{l_i=0}^{k_i}{k_i\choose l_i}
\nonumber\\&&\hspace{-180pt}\times
{(-1)^{\sum_{j=1}^n l_j}
\over (1+\sum_{j=1}^n l_j)n^{\sum_{j=1}^n l_j}}
\biggr\lbrace
\Bigl( \sum_{j\ne i}\dot G_{ij} +1 \Bigr)
^{1+\sum_{j=1}^n l_j}
- (-1)^{k_i-l_i}
\Bigl(
\sum_{j\ne i}\dot G_{ij} -1
\Bigr)^{1+\sum_{j=1}^n l_j }
\biggr\rbrace
\, .
\nonumber\\
\non
\ear
For example, all integrals appearing in the application of the worldline formalism
to the calculation of the heat-kernel (= large mass) expansion in scalar or spinor QED
can, after suitable IBP's  \cite{5}, be performed by a recursive application of this formula. 

Very useful are also the following ``chain integral' formulas, which are all that is needed for
calculating the scalar and spinor QED one-loop $N$-photon amplitudes in the low-energy limit \cite{51}:
\begin{eqnarray}
\int_0^1 du_2\ldots du_n
\dot G_{12}\dot G_{23}\ldots\dot G_{n(n+1)} 
&=& -{2^n\over n!}B_n(\vert u_1-u_{n+1}\vert)
{\rm sign}^n(u_1-u_{n+1})
\, ,
\label{bernoulli}\\
\int_0^1 du_2\ldots du_n
{G_F}_{12}{G_F}_{23}\ldots{G_F}_{n(n+1)} 
&=& {2^{n-1}\over{(n-1)!}}
E_{n-1}(\vert u_1-u_{n+1}\vert)
{\rm sign}^n(u_1-u_{n+1})
\label{euler}
\end{eqnarray}
\no
(the $B_n(x), E_n(x)$  are the Bernoulli and Euler polynomials.) 

\section{The magic magnetic master integral}

Even the worldline integrals for the low-energy limits of the $N$-photon amplitudes in the presence of an external constant magnetic field
can still be solved in general \cite{156,lopezlopez,ahcoed}. As we have seen in section \ref{sec:external}, adding on the
magnetic field changes $\dot G$ to $\dot{\cal G}_B$ which involves the three component functions $\dot G$, $S_B$ and $A_B$.  
Introducing the function
\bear
H_{ij} (z) \equiv 
\frac{e^{z \dot G_{ij}}}{\sinh z} - \frac{1}{z} 
\label{defH}
\ear
these three functions can be written as 
\bear
\dot G_{ij} &=& H_{ij}(0) \, , \quad
S_{Bij}(z) =\half \Bigl\lbrack H_{ij}(z) + H_{ij}(-z) \Bigr\rbrack \, , \quad
A_{Bij}(z) = \half \Bigl\lbrack H_{ij}(z) - H_{ij}(-z) \Bigr\rbrack
\nonumber\\
\ear
so that it now boils down to computing chain integrals of the function $H_{ij} (z)$ instead of \eqref{bernoulli}. 
And the nice thing about this function is that it self-reproduces under folding in the following, fully permutation symmetric way:
\bear
H_{ik}^{(2)}(z,z') &\equiv &
\int_0^Td\tau_j H_{ij}(z) H_{jk}(z') = 
\frac{H_{ik}(z)}{z'-z} + \frac{H_{ik}(z')}{z-z'}
\, ,
\nonumber\\
H_{il}^{(3)}(z,z',z'') &\equiv &
\int_0^Td\tau_j \int_0^T d\tau_k H_{ij}(z) H_{jk}(z') H_{kl}(z'') 
\nonumber\\
&=&
\frac{H_{il}(z)}{(z'-z)(z''-z)}
+\frac{H_{il}(z')}{(z-z')(z''-z')}
+\frac{H_{il}(z'')}{(z-z'')(z'-z'')}
\, ,
\nonumber\\
&\vdots &\nonumber \\
H^{(n)}_{i_1i_{n+1}}(z_1,\ldots,z_n) & =&  \sum_{k=1}^n \frac{H_{i_1i_{n+1}}(z_k)}{\prod_{l \ne k} (z_l - z_k)}
\, .
\label{magic}
\ear 
In the spinor QED case, one needs also an analogous generalization of \eqref{euler}, which turns out \cite{156} to be exactly the same equation \eqref{magic}
just with $H_{ij} (z)$ replaced by $H^F_{ij} (z) \equiv G_{Fij} \frac{e^{z \dot G_{ij}}}{\cosh z}$.  

\section{One-loop four-point amplitudes}

Next, let us look at the computation of one-loop four-point amplitudes at finite energy, which presents the next level of difficulty 
in this integration problem.  
For example, the four-photon amplitude in spinor QED can, using the master formula \eqref{qedmaster}, suitable IBP's \cite{41},
tensor decomposition and the Bern-Kosower replacement rules mentioned above, be written very compactly in terms of the following
five parameter integrals \cite{136,137}

\bear
	\Gamma^{(k)}_{\cdots}
	&=& 
	\int_0^\infty \frac{dT}{T} T^{4-\frac{D}{2}}\e^{-m^2T}
	\int_0^1\prod_{i=1}^4du_i\, \Gamma^{(k)}_{\ldots}(\Gd_{ij})\,
\e^{-\Lambda T}
		\label{gamma}
	\ear
	($k=1,\ldots,5$),
	where $s,t,u$ are the Mandelstam variables, 
	\bear
\Lambda = \frac{1}{2} \bigl\lbrack (G_{12} + G_{34}) s + (G_{13}+G_{24})t + (G_{14}+G_{23}) u \bigr\rbrack
\ear
and
\bear
\Gamma^{(1)}_{(1234)} &=& \Gd_{12}\Gd_{23}\Gd_{34}\Gd_{41} - G_{F12}G_{F23}G_{F34}G_{F41} \,, \\
\Gamma^{(2)}_{(12)(34)} &=& \bigl(\Gd_{12}\Gd_{21} - G_{F12}G_{F21}\bigr)  \bigl(\Gd_{34}\Gd_{43} - G_{F34}G_{F43}\bigr) \,, \\
\Gamma^{(3)}_{(123)1} &=& \bigl(\Gd_{12}\Gd_{23}\Gd_{31} - G_{F12}G_{F23}G_{F31}\bigr) \Gd_{41} \,, \\
\Gamma^{(4)}_{(12)11} &=&  \bigl(\Gd_{12}\Gd_{21} - G_{F12}G_{F21}\bigr)  \Gd_{13}\Gd_{41} \,, \\
\Gamma^{(5)}_{(12)12} &=&  \bigl(\Gd_{12}\Gd_{21} - G_{F12}G_{F21}\bigr)  \Gd_{13}\Gd_{42} \, . 
\label{hatgamma}
\ear
The integrals contain the contributions of the well-known three inequivalent diagrams for light-by-light scattering,
and the challenge is once more to perform the integrations without returning to ordered sectors. 
This can be achieved \cite{victor,inprep} using IBP and the following formula, that can be used to integrate out
one of the photon legs and express the result in a way that is still valid for any ordering of the remaining three photons,
\bear
 \int_0^1 du_4\,\e^{-\Lambda T} &=& 
 \frac{1}{T}
\biggl\lbrack \frac{2}{u+\dot G_{12}t+\dot G_{13}s}+\frac{2}{u-\dot G_{12}t-\dot G_{13}s}\biggr\rbrack 
\,\e^{\half(G_{12}+G_{13}-G_{23})uT} 
\nonumber\\&&
+ \frac{1}{T}
\biggl\lbrack \frac{2}{t+\dot G_{23}s+\dot G_{21}u}+\frac{2}{t-\dot G_{23}s - \dot G_{21}u}\biggr\rbrack 
\,\e^{\half(G_{12}+G_{23}-G_{13})tT} 
\nonumber\\&&
+ \frac{1}{T}
\biggl\lbrack \frac{2}{s+\dot G_{31}u+\dot G_{32}t} +
\frac{2}{s-\dot G_{31}u-\dot G_{32}t} 
  \biggr\rbrack 
\,\e^{\half (G_{13}+G_{23}-G_{12})sT} 
\, .
\nonumber\\
\label{app-fullintG}
\ear

\section{Integration-by-parts algorithms}

Fortunately, in many cases solving this hard and non-standard problem is not really necessary, since just having the contributions 
of many Feynman diagrams contributing to an amplitude combined in one big integral can be suggestive of certain integration-by-parts (`IBP') procedures that can already lead to substantial simplifications.
A good example is \cite{15}, where the two-loop $\beta$-function coefficient for spinor QED was obtained using an IBP that led to extensive cancellations,
leaving only trivial integrals to compute.  
Usually those IBP's are in the $\tau_i$ variables, including the algorithm that is the basis of the
Bern-Kosower rules mentioned above. There the objective is to remove all the second derivatives $\ddot G_{ij}$ occurring in the
master formula \eqref{qcdmaster}, which can be done for any $N$, and preserving the full permutation symmetry of the master formula \cite{26}. 
It leads to a decomposition of the integrand in terms of ``cycles'' and ``tails'' \cite{strassler2,26} that, apart from facilitating the application of the
Bern-Kosower rules, more recently has also been found very useful for the extraction of Berends-Giele currents in BCJ gauge \cite{140,142}.
Applied to the off-shell amplitudes it leads to the automatic appearance of gauge-invariant structures, both in the abelian \cite{91} and non-abelian \cite{105} case.

In the forthcoming \cite{inprep}, where we present a complete recalculation of the scalar and spinor QED (on-shell) four-photon amplitudes using the worldline formalism,
we show that even IBP's in the global proper-time parameter $T$ can sometimes be useful. 
Namely, the (helicity-factor corrected) difference between the ``all plus'' and ``one minus'' amplitudes in the scalar QED case can, starting from the master formula \eqref{qedmaster} and using the
symmetries of the universal exponent $\Lambda$, be massaged into the following form,
\bear
\frac{\Gamma^{++++}}{C^{++++}} &=& \frac{\Gamma^{-+++}}{C^{-+++}} 
 - \frac{8}{st} 
  \int_0^\infty dT \e^{-m^2T}  \prod_{i=1}^4 \int_0^1 du_i
\Bigl\lbrace
- 3 + \sum_{i<j=1}^4 \delta(u_i-u_j)
\Bigr\rbrace
\frac
{\partial}{\partial T}
\e^{-\Lambda T}
\ear
where 
$C^{++++}=\dfrac{t}{4s} [12]^2 [34]^2$ and 
$C^{-+++} = - \dfrac{1}{4u} \left\langle 13 \right\rangle^2 [23]^2 [34]^2$.
In the massless case this integral collapes to a boundary term at $T=0$, 
while in the massive case it allows one to write the difference in terms of 
scalar box and triangle functions after a single IBP in T.

\section{Two-loop QED vacuum polarization tensors}

\begin{figure}[htbp]
\begin{center}
\includegraphics[scale=0.45]{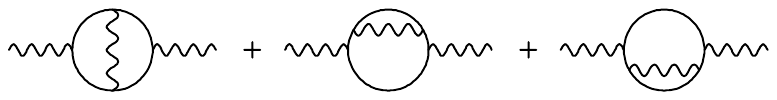}
\vspace{-240pt}
\caption{Two-loop spinor QED vacuum polarisation diagrams.}
\label{fig-2loopvpdiag}
\end{center}
\end{figure}

All the master formulas mentioned above are valid off-shell, and therefore  can be used as building blocks for the construction of multi-loop amplitudes. 
For example, from the four-photon amplitudes we can construct the two-loop photon vacuum polarizations in scalar and spinor QED, depicted
in Fig. \ref{fig-2loopvpdiag} (for the spinor QED case). 

\vspace{-100pt}

Similarly, the three-loop quenched vacuum polarization amplitudes can be constructed from the off-shell one-loop six-photon amplitudes.
This type of sums of diagrams is known to suffer from particularly extensive cancellations (see, e.g., \cite{brdekr}), so that having formulas that combine all of them
into one big integral ought to be relevant. 

At the two-loop level, the worldline representation has been used for the already mentioned calculation of the $\beta$-function coefficients \cite{15},
and also for studying the first radiative corrections to the Euler-Heisenberg and Weisskopf Lagrangians \cite{18,40,41,51}. The logical next
step is a calculation of the two-loop vacuum polarization tensor for general momentum, and this is what two of the present authors (V. B. and C.S.) will present in a
forthcoming publication (for the scalar QED case). 

In conventions where the photons 3 and 4 are the ones that get sewn together 
by an internal propagator, the integrand of the two-loop vacuum polarisation gets naturally
expressed in terms of the variables $G_{12},G_{34}$ and $ C_{12;34} = G_{13} - G_{14} - G_{23} + G_{24}$ \cite{160}. The prototypical integral to be performed in this
two-loop calculation is 
\bear
I^{m}_{k} \equiv \int_0^1du_3\int_0^1du_4 \frac{C_{12;34}^{2m}}{G_{34}^k}
\ear
$(1\leq k\leq m)$ which is already much more challenging than anything we have seen above at the one-loop level. 
The result of this integration should depend on the remaining variables $u_1,u_2$ only through the function $G_{12}$,
or equivalently (since $\dot G_{12}^2 = 1 - 4 G_{12})$ through $\dot G_{12}$. 
Curiously, we find that, for arriving at a relatively simple form of the
result, both $G_{12}$ and $\dot G_{12}$ should be used:
\bear
I_k^m &=& 
\gamma_k^m 
\biggl\lbrace
 \ln (G_{12}) \frac{(1-\dot G_{12})^{2m}}{4^m}\Bigl\lbrack  b_+(k,m)(1-\dot G_{12})
+  b_-(k,m) (1+\dot G_{12})
\Bigr]\nonumber\\ 
&& \!\!\!
+ 2 \sum_{j=0}^m \Bigl(b_+(k,m)d_j^m + b_-(k,m)d_{j-1}^{m-1}\Bigr)
\Bigl\lbrack
\dot G_{12} \ln \Bigl(\frac{1+\dot G_{12}}{2}\Bigr)
- 
 \sum_{n=1}^{2m-k-j}\alpha_nG_{12}^n\Bigr\rbrack 
 G_{12}^j 
\biggr\rbrace
\, ,
\nonumber\\
\label{taylorIkm}
\ear

\bear
b_+(k,m) &=& - (2m+1-k)(m-1/2)\, , \quad
b_-(k,m) = -(k-1)(m+1/2) 
\, ,
\ear
\bear
c_j^m &=&{ \binom{2m-j+1}{j} }\, , \quad
%
%
%
d_j^m = (-1)^j (c_j^m -c_{j-1}^{m-1})
\, ,
\ear

\bear
\gamma_k^m &=& 
\frac{ (-1)^k 4^m}{(k-1)!(m+\half)}
\frac{(2m-k)!}{(2m-2k+2)!}
\, ,
\\
\alpha_n&=& C_{n-1} \bigl(\psi(n-1/2) - \psi(n+1) + \ln 4\bigr) 
\, .
\ear

Here the $C_n$ are the Catalan numbers, and $\psi(x)$ is the digamma function. 

\section{Three-loop $\beta$-function coefficients in $\phi^4$ theory}

\begin{figure}[htbp]
\begin{center}
 \includegraphics[width=0.45\textwidth]{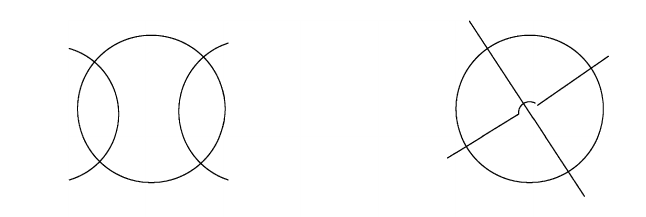}
\caption{Planar (P) and nonplanar (NP) diagram contributing to the three-loop $\beta$-function in $\phi^4$ theory.}
\label{fig-beta}
\end{center}
\end{figure}

Connecting also the remaining photon legs 1 and 2 with a propagator, we can
make further use of the one-loop four-photon amplitudes to construct the three-loop vacuum
diagrams in QED. The prototypical worldline integral to compute then becomes
\bear
I^m_{kl}  \equiv \int_0^1 du_1 du_2du_3du_4 \frac{C_{12;34}^{2m}}{G_{12}^l G_{34}^k}
\label{defImkl}
\ear
($1\leq k,l\leq m$). 
With \eqref{taylorIkm} in hand, it is possible to show that
\bear
I^m_{kl} &=&
4^{m+1} 
\frac
{(\Gamma(2m-k-l+3))^2}
{\Gamma(4m-2k-2l+5)}
\frac{1}
{(2m-l+1)(2m-l+2)}
\nonumber\\ && \times
\biggl\lbrace
\frac{2m-2l+3}{2m-k+1}
\,{}_3F_2(1,k,l-1;2m-k+2,2m-l+3;1)\nonumber \\
&& - \frac{2m(2m-2l+1)}{(2m+1)(2m-k+2)}\,
\,{}_3F_2(1,k,l;2m-k+3,2m-l+3;1)
\biggr\rbrace
\, .
\label{ImklF}
\ear
Since vacuum diagrams are of little physical interest, let us now switch to scalar
$\phi^4$ theory, where such three-loop vacuum energy diagrams can occur as the
zero-momentum limit of more relevant three-loop four-point diagrams (Fig. \ref{fig-beta}), from which one can extract $\beta$-function coefficients.

 In \cite{41} one of the present authors (C.S.)
showed that in the worldline formalism the calculation of the sum of the $1/\epsilon$ poles
of these diagrams can be reduced to the calculation of the single integral
\begin{eqnarray}
I_{\rm reg}&=& 
\int\limits_{\hspace{4mm}1234}
\biggl[-{4\over C_{12;34}^2}{\rm ln}
\Bigl(1-{C_{12;34}^2\over 4G_{12}G_{34}}\Bigr)
-{1\over G_{12}G_{34}}\biggr]
\, .
\label{Ireg}
\end{eqnarray}
However, the actual calculation of this integral was done by decomposing it into its planar (non-crossing propagators) and
non-planar (crossing propagators) sectors $I_{\rm reg}^P$ and $I_{\rm reg}^{NP}$ , resulting in
\bear
I_{\rm reg}^P = 4\zeta(2)-4 \, ,\quad  I_{\rm reg}^{NP} = 12\zeta(3)  -8\zeta (2) \, .
\ear
Using the above formulas, this decomposition can now be avoided. Expanding the logarithm in \eqref{Ireg} we get
\bear
I_{\rm reg}&=& 
\sum_{n=2}^{\infty} \frac{I^{n-1}_{nn}}{n\, 4^{n-1}}
\, .
\label{Iregsum}
\end{eqnarray}
From \eqref{ImklF} one finds the special case
\bear
I^{n-1}_{nn} &=& 4^n
\Bigl\lbrack
\frac{1}{(n-1)^2} 
+\frac{ \psi'(n)}{n-\half}
\Bigr\rbrack
\ear
which reduces this more ``principled'' recalculation of $I_{\rm reg}$ to two easy summation problems, 
\bear
\sum_{n=1}^\infty \frac{1}{n^2 (n+1)} = \zeta(2)-1 \, , \quad 
\sum_{n=1}^\infty \frac{\psi'(n+1)}{(n+\half) (n+1)} = 3\zeta(3) -2\zeta(2) \, .
\ear

\section{Conclusions}

We have summarized the state of a long-term effort to exploit the potential of the worldline formalism to tame the
proliferation of Feynman diagrams in QED multiloop calculations such as of the
anomalous magnetic moment. 
This requires the development of analytical techniques for
the global calculation of integrals involving the worldline Green's functions, and the main progress reported here is that such 
methods are now available for the topology of two-loop self-energy and three-loop vacuum diagrams. We have also shown
an unusual example of relating different helicity amplitudes by a total derivative in the global proper-time, which looks promising for
generalizations.


\begin{thebibliography}{99}

\bibitem{feynman1950}
R. P. Feynman, Phys. Rev. {\bf 80} (1950) 440.

\bibitem{feynman1951}
R. P. Feynman, Phys. Rev. {\bf 84} (1951) 108.


\bibitem{fradkin66}
E. S. Fradkin, Nucl. Phys. {\bf 76} (1966) 588.

\bibitem{fragit91}
E.S. Fradkin, D.M. Gitman,
Phys. Rev. D {\bf 44} (1991) 3230.

\bibitem{dashsu}
K. Daikouji, M. Shino, Y. Sumino, Phys. Rev. D {\bf 53} (1996) 4598, hep-ph/9508377. 

\bibitem{18}
M. Reuter, M. G. Schmidt and C. Schubert, 
Ann. Phys. (N.Y.) {\bf 259} (1997) 313, hep-th/9610191. 

\bibitem{130}
N. Ahmadiniaz, V.M. Banda Guzm\'an, F. Bastianelli, O. Corradini, J.P. Edwards and C. Schubert,
JHEP {\bf 2008} (2020) 049, arXiv:2004.01391 [hep-th].

\bibitem{131}
N. Ahmadiniaz, V.M. Banda Guzm\'an, F. Bastianelli, O. Corradini, J.P. Edwards and C. Schubert,
JHEP {\bf 01} (2022) 050, arXiv:2107.00199 [hep-th].

\bibitem{polbook}
A.M. Polyakov, {\sl Gauge Fields and Strings}, Harwood 1987.

\bibitem{strassler}
M. J. Strassler, 
Nucl. Phys. B {\bf 385} (1992) 145, hep-ph/9205205.

\bibitem{41}
C. Schubert, 
Phys. Rept. {\bf 355}, 73 (2001), arXiv:hep-th/0101036.  

\bibitem{shaisultanov}
R.~Zh. Shaisultanov, 
Phys. Lett. B {\bf 378} (1996) 354, hep-th/9512142.

\bibitem{ditsha}
W. Dittrich, R. Shaisultanov, Phys. Rev. {\bf D 62} (2000) 045024, hep-th/0001171.

\bibitem{40} 
C. Schubert,
Nucl. Phys. B {\bf 585}, 407 (2000), hep-ph/0001288.

\bibitem{17}
S.L. Adler and  C. Schubert, 
Phys. Rev. Lett. {\bf 77} (1996) 1695, hep-th/9605035.

\bibitem{156}
N. Ahmadiniaz, M. A. Lopez-Lopez and C. Schubert,
Phys. Lett. B {\bf 852} (2024) 138610, arXiv: 2312.07047 [hep-th].

\bibitem{lopezlopez}
M.A. Lopez Lopez, Phys. Lett. {\bf B} 860 (2025) 139157, hep-th/2408.16474. 

\bibitem{ahcoed}
I. Ahumada, P. Copinger and J.P. Edwards,
hep-th/2507.22308.

\bibitem{51}
G. V. Dunne and C. Schubert, 
JHEP {\bf 0208}, 053 (2002), hep-th/0205004. 


\bibitem{61}
F. Bastianelli and C. Schubert,
JHEP {\bf 0502}, 069 (2005), gr-qc/0412095.  

\bibitem{71}
F. Bastianelli, U. Nucamendi, C. Schubert, and V. M. Villanueva, 
JHEP {\bf 0711} (2007) 099, arXiv:0710.5572.
 
\bibitem{161}
N. Ahmadiniaz, F. Bastianelli,  F. Karbstein and C. Schubert, 
arXiv:2601.23279 [hep-th].

\bibitem{141}
James P. Edwards and C. Schubert,
Phys. Lett. B {\bf 22} (2021) 136696, e-Print: 2105.08173 [hep-th]. 

\bibitem{ceir}
P. Copinger, J.P. Edwards, A. Ilderton and K. Rajeev,
Phys. Rev. {\bf D} 109 (2024) 065003, arXiv:2311.14638 [hep-th].


\bibitem{berkosNPB}
Z. Bern and D.A. Kosower, 
Nucl. Phys. B {\bf 379} (1992) 451.

\bibitem{158}
F. Bastianelli, O. Corradini, J.P. Edwards, D.G.C. McKeon and C. Schubert,
JHEP {\bf 11} (2024) 152, arXiv:2406.19988 [hep-th].

\bibitem{bafemi}
F. Bastianelli, F. Fecit and A. Miccichè, 
JHEP 09 (2025) 201, arXiv:2507.15943 [hep-th]. 

\bibitem{15} 
M.G. Schmidt and C. Schubert, 
Phys. Rev. D {\bf 53}, 2150 (1996), hep-th/9410100.

\bibitem{135}
J.~P.~Edwards, C.~M.~Mata, U.~M\"uller and C. Schubert,
SIGMA {\bf 17} (2021), 065, arXiv:2106.1207 [hep-th]. 

\bibitem{5} 
M.G. Schmidt and C. Schubert, 
Phys. Lett. B {\bf 318}, 438 (1993), hep-th/9309055. 

\bibitem{136}
N. Ahmadiniaz, C. Lopez-Arcos, M. A. Lopez-Lopez and C. Schubert,
Nucl. Phys. B {\bf 991} (2023) 116216; arXiv:2012:11791 [hep-th]. 

\bibitem{137}
N. Ahmadiniaz, C. Lopez-Arcos, M. A. Lopez-Lopez and C. Schubert,
Nucl. Phys. B {\bf 991} (2023) 116217;  arXiv:2303.12072 [hep-th]. 

\bibitem{victor}
V.~M. Banda Guzmán, 
JHEP 07 (2025) 159; arXiv:2505.04157 [hep-th]. 

\bibitem{inprep}
V.~M. Banda Guzmán,  J.~P. Edwards, C.~M. Mata, L.~ A. Rodriguez Chacón and C.S., 
``Four-photon amplitudes in scalar and spinor QED'', in preparation. 

\bibitem{26}
C. Schubert,
Eur. Phys. J. C{\bf 5}, 693 (1998), hep-th/9710067. 

\bibitem{strassler2}
M.J. Strassler, 
SLAC-PUB-5978 (1992) (unpublished).

\bibitem{140}
N. Ahmadiniaz, F. M. Balli, C. Lopez-Arcos, A. Quintero Velez and C. Schubert,
Phys. Rev. D {\bf 104} (2021) L941702, arXiv:2105.06745 [hep-th]. 

\bibitem{142}
N. Ahmadiniaz, F. M. Balli, O. Corradini, C. Lopez-Arcos, A. Quintero Velez and C. Schubert, 
Nucl. Phys. B {\bf 975} (2022) 115690, arXiv:2110.04853 [hep-th].

\bibitem{91}
N. Ahmadiniaz, C. Schubert and V.M. Villanueva, 
JHEP {\bf 1301}, 312 (2013), arXiv:1211.1821 [hep-th].

\bibitem{105}
N. Ahmadiniaz and C. Schubert,
Int. J. Mod. Phys. E {\bf 25} (2016) 1642004, arXiv:1811.10780 [hep-th]. 


\bibitem{brdekr}
D. J. Broadhurst, R. Delbourgo and D. Kreimer, 
Phys. Lett. B 366 (1996) 421; hep-ph/9509296.


\bibitem{160}
N. Ahmadiniaz, V.M. Banda Guzmán, J.P. Edwards, M.A. Lopez-Lopez, C.M. Mata, L.A. Rodriguez Chacón, C. Schubert and R. Shaisultanov, 
Proc. of {\sl Workshop Loops and Legs in Quantum Field Theory, 14-19 April 2024, Wittenberg, Germany}, 
publ. in {\it Proc. of Science} (LL2024) {\bf 011}, arXiv:2407.07388 [hep-th].




\end{thebibliography}
\end{document}